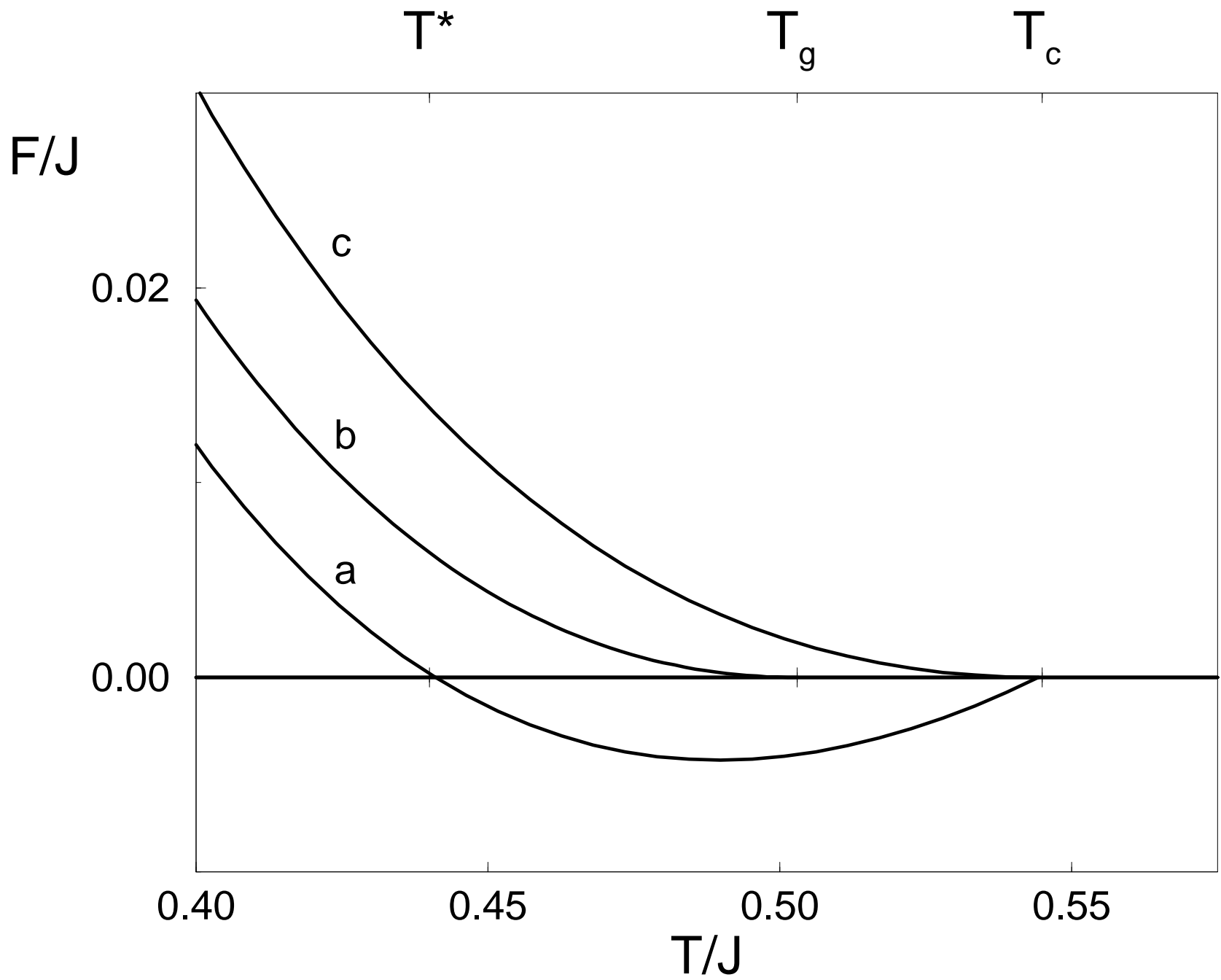

# To maximize or not to maximize the free energy of glassy systems, != ?

Th. M. Nieuwenhuizen

*Van der Waals-Zeeman Laboratorium, Universiteit van Amsterdam*
*Valckenierstraat 65, 1018 XE Amsterdam, The Netherlands*

The static free energy of glassy systems can be expressed in terms of the Parisi order parameter function. When this function has a discontinuity, the location of the step is determined by maximizing the free energy. In dynamics a transition is found at larger temperature, while the location of the step satisfies a marginality criterion. It is shown here that in a replica calculation this criterion minimizes the free energy. This leads to genuine first order phase transitions at the dynamic transition point. Though the order parameter function is the same as in the long-time limit of a dynamical analysis, thermodynamics is different.

The theoretical understanding of systems with broken ergodicity has gained a lot from the study of spin glasses, see Mezard, Parisi and Virasoro, [1] and Fischer and Hertz [2] for reviews. The solution for the celebrated Sherrington-Kirkpatrick (SK) mean field spin glass [3] was formulated by Parisi. [4] [1] The order parameter is a function $q(x)$ on the interval $0 < x < 1$. A non-constant form of $q$ implies breaking of ergodicity. In the SK-model $q$ has a continuous part followed by a plateau. Since it takes an infinity of values, replica symmetry and ergodicity are broken to infinite order. The free energy of the spin glass phase exceeds the continuation of the paramagnetic branch, as in real glasses.

Recently a spherical spin glass model was proposed by the present author, which has the same critical behavior as the SK model, but can be solved exactly in the whole low-temperature phase. [5] The same shape of the order parameter function persists up to $T = 0$. As usual for classical vector spins or spherical spins, the entropy goes to minus infinity at low $T$. A recently introduced quantum version of the spherical model [6] improves this behavior. It is still solvable exactly and yields an entropy that is non-negative and vanishes at $T = 0$.

Another canonical case, the one of interest in the present work, is that there occurs a one step replica symmetry breaking (1RSB). Then $q$ consists of two plateau's, $q(x) = q_0$ for $0 < x < x_1$ and $q(x) = q_1$ for $x_1 < x < 1$. (In our present discussion the external field will vanish, which implies $q_0 = 0$.) This situation occurs, for instance, in the "simplest spin glass", namely the Ising spin glass with $p$-spin interactions for $p \to \infty$, [7] in its spherical analog at any $p > 2$, [8] in the Potts glass, [9] in the binary perceptron, [10] [11] in a neural network with modified pseudo-inverse learning rule, [12] and in models for protein folding. [13] [14] According to the Parisi description one looks for the maximum of the free energy as function of the break point $x_1$. For $T$ below some $T_g$ one finds a glassy phase that is usually stable against small fluctuations. The free energy *exceeds* the value obtained by continuing the paramagnetic free energy. The transition is first order in the sense that the order parameter is discontinuous, but the latent heat vanishes, since, if finite, it would be negative. As such a transition takes place before the paramagnet becomes unstable with respect to small fluctuations, one has to assume that the paramagnet has some non-perturbative instability. [2]

Recently there have been dynamical studies of the binary perceptron by Horner, [15] and $p$-spin interaction spin glasses by Kirkpatrick and Thirumalai [16] and by Crisanti, Horner, and Sommers. [18] This revealed that also in dynamics a similar 1RSB transition occurs. However, it sets in at some $T_c$ that exceeds $T_g$. It was noted [16] [17] that in the long-time limit the solution satisfies a *marginality criterion*, that is to say, the lowest fluctuation eigenvalue vanishes. Indeed, it is quite natural that dynamics will get trapped in such a mode, if it can reach it. Therefore we wish to consider the marginality not as some accidental property but rather as a fundamental aspect. Using this criterion, we perform a replica analysis. We would like to think of our procedure as an "infinite-time limit". We stress that these times, though infinite, are much smaller than the ones needed to reach the *static limit*. As we shall see, the results obtained partially disagree with the ones of the *long-time limit* of the dynamical analysis.

Let us start by noting that in the replicated partition sum one should integrate over all shapes of the order parameter $q_{\alpha\beta}$. In a saddle point approach one should thus consider all solutions of the saddle point equation of the replicated free energy, i.e. $\partial F_n / \partial q_{\alpha\beta} = 0$. For a 1RSB solution the variations include variation w.r.t. to the value of the order parameter, $\partial F / \partial q_1 = 0$, as well as with respect its parametrization, i.e., $\partial F / \partial x_1 = 0$. The latter is called the *stationarity criterion*. The validity of this procedure has been confirmed in the exactly solvable " simplest spin glass". [7]

Let us now explain how the marginality condition shows up in a replica calculation. At any given break point $x_1$ one can evaluate the free energy $F(q_1(x_1); x_1)$ by solving the extremality condition $\partial F / \partial q_1 = 0$. By varying $x_1$ between 0 and 1 no local minimum is found. The minimal value of $F$ is therefore set by one of the boundaries of the allowed $x_1$ interval. The upper boundary $x_1 = 1$ is not very interesting; it corresponds to a paramagnet. Since the free energy also has to be (at least marginally) stable against small fluctuations, the lower extremal value of $x_1$ is set by equating the lowest eigenvalue of the fluctuation matrix to zero, $\Lambda_1(x_1) = 0$. This is exactly the condition satisfied automatically in the dynamical solution of most systems mentioned above, see, e.g., [8] [16] [15] [18]. We have thus found that, by definition, the marginality criterion leads to minimization of the free energy. However, as discussed below, in some models the condition $\Lambda_1 = 0$ is never satisfied and there is only the static transition.

We shall demonstrate our point on some specific models. First consider the spherical $p$-spin interaction spin glass, as studied by Crisanti and Sommers (CS). [8] For a system with $N$ spins the Hamiltonian reads

$$\mathcal{H} = \sum_{i_1 < i_2 < \cdots < i_p} J_{i_1 i_2 \cdots i_p} S_{i_1} S_{i_2} \cdots S_{i_p} \qquad (1)$$

with independent Gaussian random couplings, that have average zero and variance $J^2 p!/2N^{p-1}$. As shown by CS, the free energy for a 1RSB solution reads

$$\beta F = -\frac{\beta^2 J^2}{4} + \frac{\beta^2 J^2}{4}(1-x_1)q_1^p \qquad (2)$$
$$- \frac{1}{2x_1}\ln(1 - (1-x_1)q_1) + \frac{1-x_1}{2x_1}\ln(1-q_1)$$

where the first term describes the paramagnetic free energy. Variation with respect to $q_1$ yields

$$\frac{p\beta^2 J^2}{2} q_1^{p-1} = \frac{q_1}{(1-q_1)(1-(1-x_1)q_1)} \qquad (3)$$

The eigenvalue for fluctuations on this plateau reads

$$\Lambda_1 = -\frac{p(p-1)\beta^2 J^2}{2} q_1^{p-2} + \frac{1}{(1-q_1)^2} \qquad (4)$$

CS fixed $x_1$ by requiring $\partial F/\partial x_1 = 0$. We denote their transition temperature by $T_g$. Adopting the marginality condition $\Lambda_1 = 0$ we obtain a transition temperature $T_c = J\{p(p-2)^{p-2}/2(p-1)^{p-1}\}^{1/2}$ and $x_1 = (p-2)(1-q_1)/q_1$. In case $p=4$, $q_1$ can be solved analytically

$$q_1 = \frac{1}{2} + \frac{1}{2}\sqrt{1 - \frac{8T}{9T_c}} \qquad (5)$$

$T_c$ exceeds $T_g$ for any $p$ and has a finite limit as $p \to \infty$, whereas $T_g \sim J/\sqrt{p}$ for large $p$. Near the transition the free energy is below the continuation of its paramagnetic value and has a larger slope. Thus one expects that a genuine first order transition takes place with positive latent heat,

$$\Delta U(T_c) = \{\frac{p-1}{p-2}\ln(p-1) - \frac{2(p-1)}{p}\}T_c \qquad (6)$$

When lowering $T$ one first passes $T_g$ where the static solution sets in. In a dynamical approach nothing special will happen at this point, since one is and remains far from the static solution [18]; the same is expected for our "infinite-time" solution. Below some even lower $T^*$ the free energy of the marginally stable state exceeds the paramagnetic one. In principle one would then return to a reentrant paramagnet ($x_1 = 1$). This phase, though stable against small fluctuations up to $T = 0$, cannot be the physical state at $T = 0$; it would lead to an internal energy that diverges as $-1/T$ for small $T$. This paradox can again be "solved" by assuming that states with free energy below the one of the marginally stable state have a non-perturbative instability.

When comparing to the results from dynamics, [18] we see that the same order parameter function is obtained: *the magnetic properties are the same as in dynamics.* This is no accident; it occurs since we adopted the marginality criterion. However, *thermodynamics is different.* Indeed, eq. (2) yields the internal energy

$$U = U_{dyn} - \Delta U = \frac{\beta J^2}{2}[-1 + (1-x_1)q_1^p] - \Delta U \qquad (7)$$

The first term is the energy occurring in dynamics. The term $\Delta U$, proportional to $dx_1/dT$, would cancel if $x_1$ were determined by a stationarity condition. For the marginality condition that is not the case; at $T_c$ $\Delta U$ yields the latent heat (6). Crisanti, Horner, and Sommers [18] define the entropy by integrating $dS \equiv (C/T)dT$ from some temperature in the paramagnet, and obtain in this way the dynamical free energy. The result, rederived in an approach via the TAP equations, [19] reads

$$\beta F_{dyn} = -\frac{\beta^2 J^2}{4}[1 - 3(p-1)q_1^p + (3p-4)q_1^{p-1}]$$
$$+ \frac{p-2}{p} - \frac{1}{2}\ln(p-1)(1-q_1) \qquad (8)$$

Whereas our "infinite-time" free energy lies below the continuation of the paramagnetic value, the long-time limit of the dynamical free energy exceeds it. It is intriguing that the finite latent heat, derived by us, is not observed in dynamics. This unusual aspect of the first order transition is probably due to the divergent time scale $1/\Lambda_1$.

In Figure 1 we present the various free energies (after subtraction of the paramagnetic value) for the model with random quartet couplings ($p = 4$).

As second example we consider a spherical model with random pair and random quartet interactions having variances $\langle J_{ij}^2 \rangle = J_2^2/N$ and $\langle J_{ijkl}^2 \rangle = 6J_4^2/N^3$. This model was studied recently by us [5], where we also allowed for inclusion of higher multiplet couplings. When the pair interactions are strong enough, a continuous spin glass transition occurs of the same type as in the SK model. The solution was derived explicitly at all $T$. Here our interest is for weak enough pair couplings. Then, as in the above case with $p = 4$, $J_2 = 0$, and $J = J_4/\sqrt{2}$, a first order transition will occur to a phase with 1RSB. We can extend eqs. (2)-(4) to this situation for spins subject to the generalized spherical constraint $(1/N) \sum S_i^2 = \sigma$. This yields that the 1RSB transition point coincides with the continuous transition at $T_c = J_2\sigma$ when $J_2 = J_4\sigma/2$. When $J_2$ is below this value, there will be a frozen phase with 1RSB. In the quantized version of the spherical model [6] [5] the role of $\sigma$ will be taken by the self-overlap $q_d \equiv q_{\alpha\alpha}$. This quantity, which has to be solved self-consistently, is temperature dependent and smaller than $\sigma$; this is a quantum effect.

Let us take as next example the $p$-state Potts spin glass. The Ginzburg-Landau expansion of the $n$-fold replicated free energy was derived by Gross, Kanter and Sompolinsky (GKS) [9]

$$\beta F_n = -\frac{\tau}{2}q_{\alpha\beta}^2 - \frac{1}{6}q_{\alpha\beta}q_{\beta\gamma}q_{\gamma\alpha} - \frac{p-2}{12}q_{\alpha\beta}^3 + \frac{y}{8}q_{\alpha\beta}^4 \quad (9)$$

summed over all repeated indices $1 \leq \alpha, \beta, \gamma \leq n$, with $q_{\alpha\alpha} = 0$. For $p = 2$ one has $y = -2/3$ so that this form reduces to the relevant part of the free energy for the Ising spin glass. It holds that $y > 0$ for $p > p^* = 2.83$. The phase diagram of this system was discussed extensively by GKS. For $2 < p < p^*$ the solution has a plateau $q_0 = 0$ for $0 < x < (p-2)/2$, then is linear up to some $x_1$ and further has a plateau $q_1 \sim \tau$ up to $x = 1$. For such a solution the marginality criterion brings no new information, since it is satisfied automatically.

For $p > p^*$ there is no (increasing) continuous solution. There are now two plateau's, $q_0 = 0$, and $q_1 > 0$. In terms of $\delta = (p-4)/2$ the free energy becomes

$$\beta F = (1 - x_1)\{\frac{\tau}{2}q_1^2 + \frac{\delta - 1 + x_1}{6}q_1^3 - \frac{y}{8}q_1^4\} \quad (10)$$

Stationarity with respect to $q_1$ gives

$$\tau + \frac{\delta - 1 + x_1}{2}q_1 - \frac{y}{2}q_1^2 = 0 \quad (11)$$

GKS adopt the Parisi condition $\partial F/\partial x_1 = 0$ to fix $x_1$. This leads to

$$q_1 = \frac{2}{5y}(\delta + \sqrt{\delta^2 + 5y\tau}) \quad 1 - x_1 = -\frac{2\delta}{5} + \frac{3}{5}\sqrt{\delta^2 + 5y\tau}$$
$$\beta F = \frac{1}{6}(1 - x_1)^2 q_1^3 \quad (12)$$

Let us first consider $p < 4$, so $\delta < 0$. Close to the transition one has $q_1 \approx \tau(1 - 5\sigma)/|\delta|$, $1 - x_1 \approx |\delta|(1 + 6\sigma)$, $\beta F \approx \tau^3\{1 - 3\sigma + 21\sigma^2\}/6|\delta|$, where $\sigma = y\tau/4\delta^2$.

In our approach we need the lowest eigenvalues. For fluctuations on the $q_0$ and $q_1$ plateau's they read

$$\Lambda_0 = -\tau + (1 - x_1)q_1 \quad \Lambda_1 = -\tau - \delta q_1 + \frac{3y}{2}q_1^2 \quad (13)$$

respectively. For $p < 4$ we assume that $\Lambda_0 = 0$. We find

$$q_1 = \frac{\delta}{2y} + \frac{1}{2y}\sqrt{\delta^2 + 4y\tau} \quad 1 - x_1 = \frac{-\delta}{2} + \frac{1}{2}\sqrt{\delta^2 + 4y\tau}$$
$$\beta F = \frac{1}{24}(1 - x_1)q_1^3[\delta + 5(1 - x_1)] \quad (14)$$

so that $\Lambda_1 > 0$. Near $\tau = 0$ one has $q_1 \approx \tau(1 - 4\sigma)/|\delta|$, $1 - x_1 \approx |\delta|(1 + 4\sigma)$ and $\beta F \approx \tau^3\{1 - 3\sigma + 20\sigma^2\}/6|\delta|$. Comparing to the static GKS solution, one sees that our "infinite-time" solution is different at order $\tau^5$. Again, it has a smaller free energy. The free energy difference with respect to the state with $\Lambda_1 = 0$ is only of order $\tau^6$.

We now consider the situation where $p$ is slightly larger than 4, $p = 4 + 2\delta$ with $0 < \delta \ll 1$. Eq. (12) remains valid. As now $\delta > 0$, it is seen that the onset occurs at $\tau_g \equiv -\delta^2/9y$, where $x_1$ becomes below unity, while $q_1$ has a finite value. GKS thus predicts a non-standard first order transition with vanishing latent heat.

Let us investigate the consequences of the condition $\Lambda_1 = 0$. We now find an onset at $\tau_c = -\delta^2/8y$, occurring before $\tau_g$. The solution reads

$$q_1 = \frac{\delta}{3y} + \frac{1}{3y}\sqrt{\delta^2 + 6y\tau} \quad 1 - x_1 = -\frac{\delta}{3} + \frac{2}{3}\sqrt{\delta^2 + 6y\tau} \quad (15)$$

It can be checked that the other eigenvalues are positive. The free energy equals

$$\beta F = \frac{1}{48}(1 - x_1)q_1^3[-\delta + 7(1 - x_1)] \quad (16)$$

which clearly becomes negative at the transition. We thus find a genuine first order phase transition with a finite latent heat. At $\tau_g$ the static GKS solution sets in. The occurrence of this state has no effect at all on the dynamical solution or on our "infinite-time" solution. Below $\tau^* \equiv -4\delta^2/49y$ the free energy of the marginally stable state becomes positive. This behavior is fully analogous to $p$-spin model, see the curves (a) and (b) in Figure 1. Though the reentrant paramagnet is stable against small fluctuations up to $\tau = 0$, it cannot be the physical branch. Indeed, if it would, then at $\tau = 0$ a new 1RSB solution should set in with $\Lambda_0 = 0$. However, this one has a negative latent heat. Thus it cannot be the physical solution, and neither can its paramagnetic precursor be. Fully analogous to previous situation we must assume that we can neglect the states with free energy below the one of the marginally stable state. Yet it remains to be shown that they are indeed unstable.

We have also applied marginality criteria for finding solutions with a finite number of steps. This effort has been rather unsuccessful. For continuous solutions (infinitely many steps) the criterion is satisfied automatically.

In conclusion, we have presented new "infinite-time" solutions of the mean field equations for glassy systems with a step in the order parameter function. These solutions are extremal in the sense that some fluctuations are marginally stable. As compared to the commonly considered static solutions, obtained by extremizing the free energy with respect to the location of the break point, it was shown that the marginality criterion leads to transitions with a lower free energy (when the $q_1$ plateau sets continuously in from 0) or to transitions that set in even at a higher temperature (when $q_1 > 0$ but $x_1$ becomes below unity). In all cases considered it was seen that the commonly considered static solutions have higher free energy.

There are also models where the fluctuations are always massive. Then the marginality condition is never satisfied. Indeed, we found no application of it in the "simplest spin glass". In this problem there is no dynamical transition point above the static one. This is in full accord with the fact that $T_c$ diverges in the $p$-spin interaction Ising spin glass. [20]

It was found that our "infinite-time" replica calculation gives the same order parameter function as dynamics in the long-time limit. However, the thermodynamical aspects are different. Most striking is that, in contradiction to the dynamical result, the free energy of the glassy state lies below the one of paramagnet. This discrepancy remains to be understood.

## ACKNOWLEDGMENTS

The author is grateful for an invitation by the W.E. Heraeus Stiftung to a seminar on neural nets, spin glasses and glasses in Bad Honnef, where discussions with J.H. Sommers and H. Horner led to the perception of the problem. Also discussions with C. de Dominicis, M. Mézard, C.N.A. van Duin, and M.C.W. van Rossum are gratefully acknowledged. This work was made possible by the Royal Dutch Academy of Arts and Sciences (KNAW).

Note added:
The difference between the free energy derived by us and its dynamical expression can be traced back to the value of the complexity.

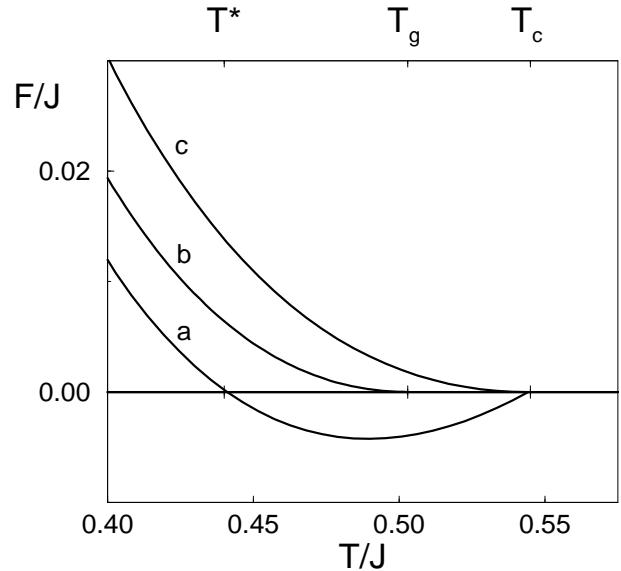

FIG. 1. Free energy of a spherical spin glass with random quartet couplings, after subtraction of the paramagnetic value. a) "Infinite-time" solution; it sets in at $T_c$. Its free energy exceeds the paramagnetic one below $T^*$. b) Static solution; it sets in at $T_g$. c) Long-time limit of the dynamical solution; it also sets in at $T_c$.